\documentstyle[11pt,aaspp]{article}
\def\t0{\theta_{\circ}}

\def\be{\begin{equation}}
\def\en{\end{equation}}

\begin{document}

\title {Mid-infrared imaging of the young binary star Hen 3-600:\\
Evidence for a dust disk around the primary}
\author{Ray Jayawardhana\altaffilmark{1,3},
Lee Hartmann\altaffilmark{1},
Giovanni Fazio\altaffilmark{1},\\
R. Scott Fisher\altaffilmark{2,3},
Charles M. Telesco\altaffilmark{2,3}
and Robert K. Pi\~na\altaffilmark{2}}
\altaffiltext{1}{Harvard-Smithsonian Center for Astrophysics, 60 Garden St., Cambridge, MA 02138; Electronic mail: rayjay@cfa.harvard.edu}
\altaffiltext{2}{Department of Astronomy, University of Florida, Gainesville, FL 32611}
\altaffiltext{3} {Visiting Astronomer, Cerro Tololo Interamerican Observatory,
National Optical Astronomy Observatories, which is operated by the Association
of Universities for Research in Astronomy, Inc. (AURA) under cooperative
agreement with the National Science Foundation.}

\begin{abstract}
We present high-resolution mid-infrared observations of the nearby late-type
young binary system Hen 3-600. The binary, at a distance of $\sim$ 50 pc,
could be a member of the TW Hydrae Association, the nearest
known group of young stars, with an age of a few million years. 
Our images make it possible for the first time to determine which star
in the pair, separated by 1.4'', harbors the mid-infrared excess detected
by IRAS. In the near-infrared, where the radiation is primarily photospheric,
Hen 3-600A (M3) and Hen 3-600B (M3.5) have a flux ratio of 1.6. At 4.8$\mu$m, 
10.8$\mu$m, and 18.2$\mu$m, the primary becomes increasingly dominant over 
the secondary, suggesting that most of the circumstellar dust in the system 
resides around Hen 3-600A. Comparison of the spectral energy distribution (SED)
of Hen 3-600A to the median SED of classical T Tauri stars suggests
that its disk may be truncated by the secondary and provides tentative
evidence for a central disk hole. The distribution of dust in the Hen 3-600
system may provide important clues to the formation and evolution of
protoplanetary disks in close binaries.
\end{abstract}

\keywords{Accretion, accretion disks -- Stars: Circumstellar Matter --
Stars: Binaries -- Stars: Formation -- Stars: Pre-Main Sequence}

\section{Introduction}
Over the past decade, infrared and millimeter observations have provided 
strong support for the single star plus accretion disk model of star 
formation (Shu, Adams, \& Lizano 1987; also see Hartmann 1998). The
disk contains most of the system's angular momentum and may help accrete
the bulk of the central star's mass. It is also the likely site of planet 
formation.

This scenario must be more complicated
in most systems because a majority of young, low-mass stars are members of 
binary systems (Ghez et al. 1993, 1997; Leinert et al. 1993; Reipurth \& 
Zinnecker 1993; Simon et al. 1995). A young binary system may contain up
to three distinct disks: circum{\it stellar} disks around one or both stars, 
with disk radii smaller than the binary's periastron separation, and a
circum{\it binary} disk lying entirely outside the binary orbit (cf. 
Jensen \& Mathieu 1997). These disks are remnants of star birth and thus
may provide clues to binary formation mechanisms. Furthermore, the stars
and the disk(s) in a binary influence each component's evolution through
dynamical interactions (e.g., Lin \& Papaloizou 1993; Artymowicz \& Lubow
1994, 1996). Therefore, the study of disks in binary systems is crucial
to our understanding of both star and planet formation processes. 

Here we report high-resolution mid-infrared imaging observations of the
nearby, young, late-type binary Hen 3-600. It was first identified by
de la Reza et al. (1989) as a T Tauri pair in the vicinity of another 
isolated young star TW Hydrae based on the presence of H$\alpha$ emission
and the Li I 6708\AA~ absorption line. Its coincidence with the IRAS source
1108-3715 was interpreted as being due to thermal dust emission associated
with one or both stars in the binary. More recently, several groups have
suggested that Hen 3-600, TW Hya and a number of other active, young stars
in the same region of the sky form a physical association with an age of
$\sim$10 Myr (Gregorio-Hetem et al. 1992; Kastner et al. 1997; Soderblom
et al. 1998; Webb et al. 1999; see Jensen, Cohen, \& Neuh\"auser 1998 for
a different viewpoint). At a distance of $\sim$50 pc, the ``TW Hydrae
Association'' is the nearest known group of young stars, and is ideally
suited to studies of circumstellar disk evolution. 

The primary goal of our observations was to determine whether the 
thermal dust emission in the Hen 3-600 system originates from one or both
stars in the binary. A secondary motivation was to place useful
constraints on the nature of the dust disk(s) by comparing the near- to
far-infrared spectral energy distribution (SED) with observations of
T Tauri disks in Taurus.

\section{Observations and Results}
Hen 3-600 was observed on February 23, 1999 with the 4-m Blanco telescope   
at Cerro Tololo Interamerican Observatory using OSCIR, the University
of Florida mid-infrared
camera. OSCIR uses a 128$\times$128 Si:As Blocked Impurity Band (BIB)
detector developed by Boeing. On the CTIO 4-m telescope, OSCIR has
a plate scale of 0.183''/pixel, which gives a field of view of 
23''$\times$23''. Our observations were made using the standard chop/nod
technique with a chopper throw of 23'' in declination. Images of Hen 3-600
were obtained in the K(2.2 $\mu$m), M(4.8$\mu$m), N(10.8 $\mu$m), and 
IHW18(18.2 $\mu$m) bands, and flux calibrated using the standard stars 
$\lambda$ Vel and $\gamma$ Cru. On-source integration times for 
Hen 3-600 were 300 seconds in K, 300 seconds in M, 600 seconds in N, and 
600 seconds in IHW18. Additional information on  OSCIR is available on the 
World-Wide Web at www.astro.ufl.edu/iag/. Our final stacked images at
each of the four wavelengths are shown in Figure 1.

In the K band, the two components of the binary are clearly resolved,
with a peak-to-peak separation of 1.4'' and a position angle of 
$215^{\circ} \pm 3^{\circ}$, consistent with previous observations.
The flux ratio we measure for the A and B components, $F_{A}/F_{B}=1.6$, 
agrees extremely well with that from Webb et al.'s (1999) K magnitudes 
based on IRTF photometry. In the M-band, the two stars are again well 
resolved, with a flux ratio of $\sim$1.9. In the N-band, only the primary 
(A) is clearly visible. Our flux 
measurement of 900$\pm$90 mJy is consistent with the IRAS 12$\mu$m flux of 
990 mJy. By subtracting an appropriately scaled PSF star from the position 
of Hen 3-600A, we are able to place an upper limit of 50 mJy on the flux
from B at N. In the IHW18 band, again only A appears on our image, with a 
total flux of 1500$\pm$150 mJy, which is consistent with the IRAS 25$\mu$m
measurement of 1750 mJy. (The errors are primarily due to uncertainties
in flux calibration.)

\section{Discussion}
\subsection{Location and nature of circumstellar dust}
Our mid-infrared images show that the IRAS-detected excess emission 
originates from Hen 3-600A, and not from Hen 3-600B, implying that
the primary has a circumstellar disk while the secondary does not.
From Keck LRIS spectra, Webb et al. (1999) measure H$\alpha$ equivalent
widths of -21.8\AA~ and -7.14\AA~ for A and B respectively. Taken together,
our mid-infrared images and Webb et al.'s spectra suggest that A is
a classical T Tauri star (CTTS) with an actively accreting disk, while B 
is a weak-line T Tauri star (WTTS) without an optically thick inner disk. 

In a spectroscopic and photometric study of 12 pre-main-sequence close
binary systems, Prato \& Simon (1997) found that in every case both
components exhibited CTTS characteristics. Thus, they suggested that the 
transition of disk optical depth, from $\tau >$1 to $\tau <$1, is roughly
coeval for both stars in a close binary and that a circumbinary envelope
could regulate the common evolution of the inner disks. Our detection of
a circumstellar disk around only the primary, and not the secondary, in
Hen 3-600 is inconsistent with that scenario. However, it does agree with
Prato (1998) who finds evidence for only circum{\it primary} disks in 4 close
binaries in a larger sample of 25 systems. We further note that in a study 
of 39 wide pre-main sequence binary systems, Hartigan, Strom, \& Strom (1994)
found that 6 systems were composed of CTTS/WTTS pairs, as judged from the 
H$\alpha$ emission (see Figure 8 of Hartigan et al. 1994). The last step in 
the evolution from the CTTS phase to the WTTS state may take place
on relatively short timescales. It is possible that Hen 3-600 is a 
transition system with one component already evolved
into a WTTS while the other is still in the process of doing so.

We plot the composite spectral energy distribution (SED) of Hen 3-600A in 
Figure 2 and compare it to the median SED for a classical T Tauri star of 
similar spectral type and to the SED of the HD 98800 quadruple system.
The solid line on the plot corresponds to the median CTTS SED for stars in 
Taurus, normalized at H (D'Alessio et al. 1999); the dashed lines show the 
quartile fluxes to provide some idea of the range of observed fluxes.
The dip at 4.8$\mu$m (M filter) in the spectrum of Hen 3-600A
is suggestive of 
a possible central hole in the disk. However, this result needs to be
confirmed with higher-precision photometry at wavelengths between
2$\mu$m--10$\mu$m. At 60$\mu$m and 100$\mu$m,
the emission from Hen 3-600A falls well below that expected for a
median $\sim$100 AU CTTS disk, implying that the disk may be significantly
smaller in radius. Assuming an optically thick disk with a temperature 
distribution $T \propto r^{-1/2}$ (Beckwith et al. 1990) suggests that 
truncating the disk at an outer radius of $\sim$25 AU would help account for 
the reduction in long-wavelength fluxes. Similar dips at 5--10$\mu$m
and again at far-infrared wavelengths are also seen in HD 98800, where
the dust is apparently associated with one of the two binary pairs
(Gehrz et al. 1999).

Theoretical calculations suggest that circumstellar disks will be truncated
by the tidal effects of a companion star in circular orbit at approximately 
0.9 of the average Roche lobe radius (Artymowicz \& Lubow 1994), which for 
roughly equal mass stars in the Hen 3-600 system would be at about a third of 
the orbital radius (Papaloizou \& Pringle 1977). Given a binary separation of 
1.4'', the disk of the primary should then be confined to $<$0.5'' angular 
radius. It is interesting to note that in another young close binary, 
L1551 IRS5 in Taurus, compact disks have been observed around each star
with the Very Large Array (VLA) at 7 mm (Rodriguez et al. 1998). In that
case, the disk radii appear to be about a quarter of the binary separation
distance. (Of course, it is possible that these outer radii are the
result of instrument sensitivity limits rather than disk truncation.)

\subsection{Age and evolutionary status of Hen  3-600}
Hen 3-600 appears to be one of several nearby young stellar systems
in the vicinity of TW Hydrae. Based on strong X-ray fluxes from all five
systems, Kastner et al. (1997) concluded that the group forms a physical 
association at a distance of $\sim$50 pc with an age of 20$\pm$10 Myr.
Webb et al. (1999) have 
identified five more T Tauri star systems in the same region of the sky
as candidate members of the ``TW Hya Association'' (TWA), based on the same
signatures of youth --namely high X-ray flux, large Li abundance, and strong
chromospheric activity-- and the same proper motion as the original
five members. Furthermore, they suggest that the wide binary HR 4796 
is also part of the Association, even though its {\it Hipparcos} 
parallactic distance of 67 pc places it further away than most other 
members of the group. The three other TWA stars with {\it Hipparcos} 
distances --TW Hya, HD 98800, and TWA 9-- are at 56, 47 and 50 pc, 
respectively. 

If all proposed members of the TWA do indeed form a coeval group with a 
common origin, they would provide an excellent sample for studying disk 
evolution timescales. In particular, using the $Hipparcos$ distance, 
and the D'Antona \& Mazzitelli (1994) evolutionary tracks, it is
possible to estimate an age of $8 \pm 3$~Myr for the M2.5 star HR 4796B 
(Jayawardhana et al. 1998).  This age is consistent with the upper bound
provided by the measurement of the strong Li absorption line at 6708 \AA~
(Stauffer et al. 1995); for this mass range, Li is predicted to be rapidly
destroyed in the stellar interior at ages $\sim 9-11$~Myr (see, e.g.,
D'Antona \& Mazzitelli 1994). We note that HR 4796B and Hen 3-600A have 
comparable Li equivalent widths and $L_{X}/L_{bol}$ ratios. Based on similar 
considerations, Soderblom et al. (1998) derive an age of 7-12 Myr for the 
HD 98800 quadruple 
system. A lower limit to the age of a few Myr is indicated 
by the isolated location of HR 4796, HD 98800, Hen 3-600, and the other TWA 
stars, since most stars of comparable or smaller ages are found in regions 
of molecular clouds and substantial interstellar dust extinction (Leisawitz, 
Bash, \& Thaddeus 1989). 

A spatially-resolved circumstellar disk was recently imaged around the
A star HR 4796A at mid-infrared wavelengths (Jayawardhana et al. 1998; 
Koerner et al. 1998; Telesco et al. 1999). The presence of an inner disk 
hole in HR 4796A, first inferred by Jura et al. (1993) on the basis of the
SED and now confirmed by mid-infrared imaging, may provide evidence for 
coagulation of dust into larger bodies on a timescale similar to that 
suggested for planet formation in the solar system (e.g., Strom et al. 1989; 
Strom, Edwards, \& Skrutskie 1993; Podosek \& Cassen 1994). If TWA stars 
are indeed coeval, it would be of significant interest to determine
whether similar disk evolution has ocurred in other systems. 

However, given that the apparent proper motions of the TWA stars are 
primarily due to solar reflex motion, it is difficult to establish a common 
origin for the candidate group members (Jensen, Cohen, \& Neuh\"auser 1998; 
Soderblom et al. 1998; Hoff, Henning \& Pfau 1998). In particular, without
a {\it Hipparcos} distance measurement, the position of Hen 3-600 on
the H-R diagram is highly uncertain. At an assumed distance of 50 pc,
Webb et al. (1999) obtain an age of a few Myr for Hen 3-600A and B from
PMS evolutionary tracks. 

In summary, it is likely that Hen 3-600 is younger than 10 Myr, based on
the presence of Li, and older than $\sim$1 Myr, given the lack of an 
associated parent molecular cloud. Possible membership of the TWA suggests
an age close to the upper end of this range while its CTTS characteristics 
point to the lower end. A more accurate age estimate may not
be possible until its distance is directly measured or a common
origin for the TWA stars is firmly established.

\subsection{Implications for disk formation and evolution in binaries}
Nearly half of the known pre-main-sequence stars in Taurus are
WTTS, i.e., they have H$\alpha$ equivalent widths $<$10 \AA~ and no
infrared excess, implying a lack of accretion disks. It has been argued
that many, if not most, WTTS have binary stellar companions which render 
disks dynamically unstable (Jensen, Mathieu, \& Fuller 1994, 1996;
Osterloh \& Beckwith 1995). Therefore, ``mixed'' binary systems like  Hen
3-600, where one star is a CTTS and the other a WTTS, present somewhat of
a puzzle. 

The exact configuration of disks --circumprimary, circumsecondary and/or 
circumbinary-- in a binary system may depend on the details of its
formation. Smoothed particle hydrodynamics (SPH) simulations by 
Bate \& Bonnell (1997) show that the fractions of infalling material that
are captured by each protostar and the fraction which forms a circumbinary
disk depend on the binary's mass ratio and the parent cloud's specific
angular momentum, $j_{infall}$. For accretion with low $j_{infall}$, most
of the infalling material is captured by the primary. For gas with intermediate
$j_{infall}$, the fraction captured by the primary decreases and
that captured by the secondary increases. For higher $j_{infall}$, more and
more gas goes into a circumbinary disk instead of circumstellar disks.
Thus, it could be that infall from a low $j_{infall}$ cloud led to a more
massive disk around the primary in Hen 3-600. However, given their roughly
equal masses, it is not clear why one star would capture much more
material than the other. One possibility is that as protostars, the two 
components had very different masses, with what is now the primary 
accumulating most of its mass at the end of accretion. Another possibility 
is that there is a very low-mass, so far undetected, close companion around 
Hen 3-600B which accelerated the depletion of its disk.

It is likely that there is not a universal evolutionary timescale for 
protoplanetary disks, especially when the influence of companion stars
is taken into account. Detailed studies of disks around a large sample 
of young binaries, including other candidate members of the TWA, may 
provide new insights to resolve this puzzle.

\bigskip
We wish to thank the staff of CTIO for their outstanding support. 
We are also grateful to Jim DeBuizer for his assistance
at the telescope. The research at CfA was supported by NASA 
grant NAG5-4282 and the Smithsonian Institution. The research at the 
University of Florida was supported by NASA, NSF, and the University of 
Florida.

\newpage

\centerline{\bf Figure Captions}

\bigskip
\bigskip

Figure 1. Final, stacked OSCIR images of Hen 3-600 in K (2.2$\mu$m)
{\it upper left}, M (4.8$\mu$m) {\it upper right}, N (10.8$\mu$m)
{\it lower left}, and IHW18 (18.2$\mu$m) {\it lower right} 
filters, smoothed with a 3-pixel Gaussian. The bars below each panel 
indicate the intensity scale. The plate scale is 0.183''/pixel.

Figure 2. (a) Composite spectral energy distribution of Hen 3-600A from near-
to far-infraed wavelengths based on measurements of Webb et al. (1999),
IRAS, and this paper. (b) The SED of HD 98800, for comparison, based on
fluxes reported by Sylvester et al. (1996). The solid line in each panel
is the median SED for Taurus CTTS, normalized at H (from D'Alessio et al. 
1999), and the dashed lines show the quartile fluxes to provide some idea 
of the range of observed fluxes.

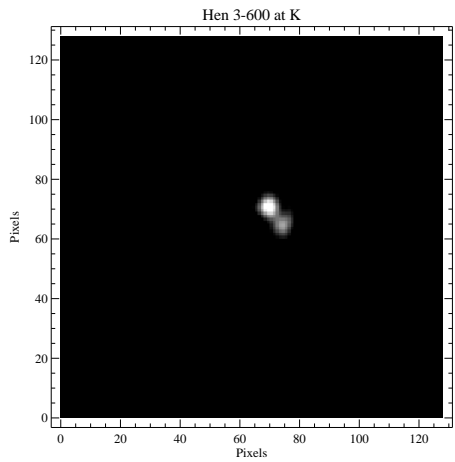
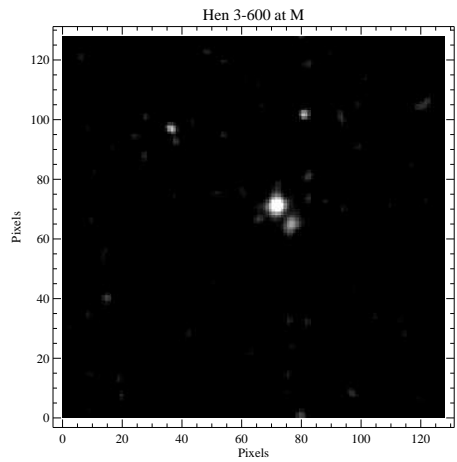
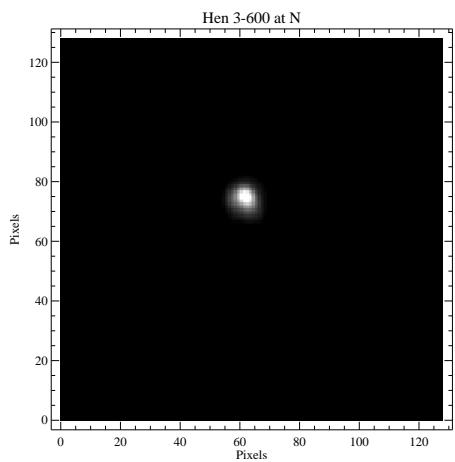
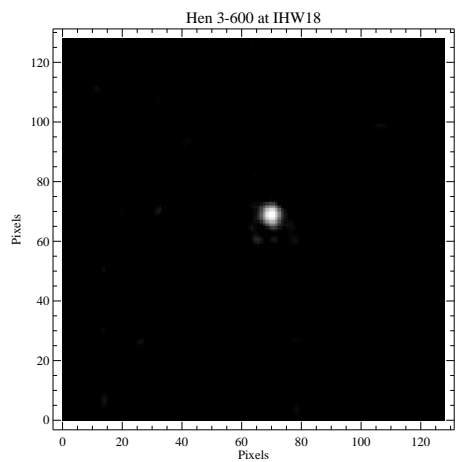
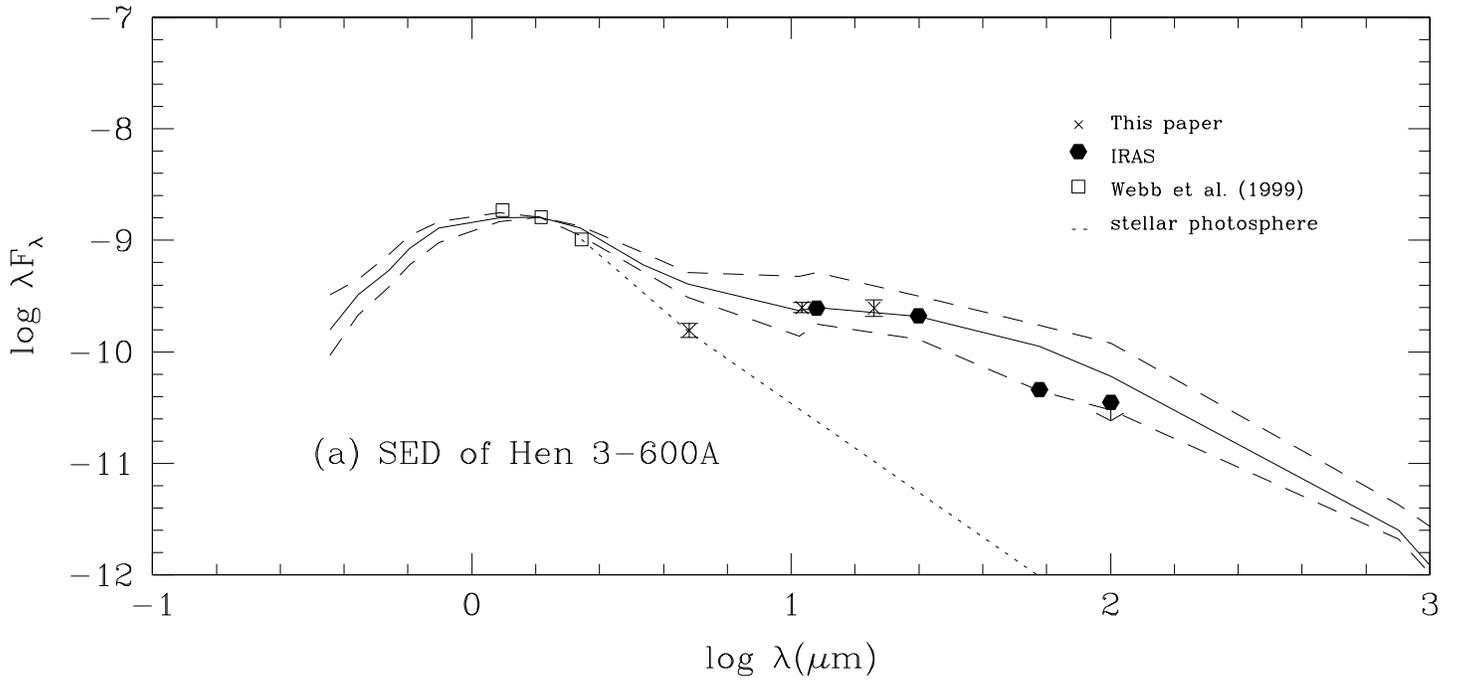
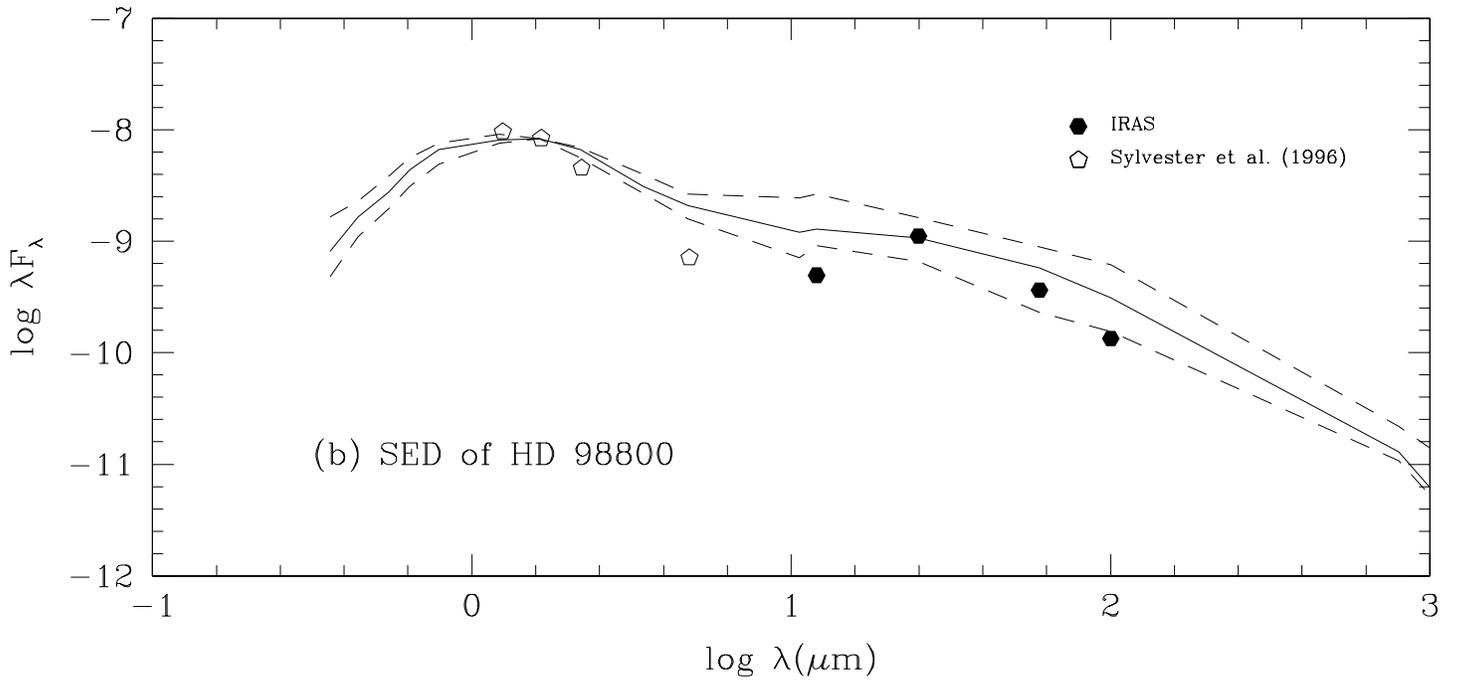
\end{document}